# Qos-based Computing Resources Partitioning between Virtual Machines in the Cloud Architecture

Evgeny Nikulchev
Moscow Technological Institute
Leninskiy pr., 38A, Moscow, Russia
119334

Evgeniy Pluzhnik
Moscow Technological Institute
Leninskiy pr., 38A, Moscow, Russia
119334

Oleg Lukyanchikov
Moscow Tecnological University MIREA
Vernadsky Avenue, 78, Moscow, Russia 119571

Dmitry Biryukov
Moscow Tecnological University MIREA
Vernadsky Avenue, 78, Moscow, Russia 119571

Elena Andrianova
Moscow Tecnological University MIREA
Vernadsky Avenue, 78, Moscow, Russia 119571

*Abstract*—Cloud services have been used very widely, but configuration of the parameters, including the efficient allocation of resources, is an important objective for the system architect. The article is devoted to solving the problem of choosing the architecture of computers based on simulation and developed program for monitoring computing resources. Techniques were developed aimed at providing the required quality of service and efficient use of resources. The article describes the monitoring program of computing resources and time efficiency of the target application functions. On the basis of this application the technique is shown and described in the experiment, designed to ensure the requirements for quality of service, by isolating one process from the others on different virtual machines inside the hypervisor.

## I. Introduction

Cloud services are one of the most popular emerging areas of modern IT industry. The development of cloud computing technology has set a number of specialized tasks requiring fundamental results. Specifically, the load control task to provide the required quality of service [1]. The allocation of resources between virtual machines and remote data centers with unknown parameters of data channels requires dynamic control of the operational parameters of virtualization and data transfer quality. Introduction of feedback from controlled parameters allows making corrective actions; on the other hand, a task of optimal control can occupy a substantial part of computing processes and communication channels [2, 3]. It is possible to get feedback from the hypervisors, on which modern cloud systems are built, and from the cloud-based software itself, if developers provided such opportunities. The idea is proposed to develop theoretical basis for the program resource management in distributed cloud infrastructures, including the methods, models, patterns and the prototype of software middleware.

Managing information systems based on full use of cloud infrastructure offers the solution for task of creating a platform that automates the allocation of resources at the lowest cost [4, 5].

Main directions:

- guaranteeing the quality of service (Quality of Service, QoS);
- optimizing resources (reduction of energy consumption, cost optimization, etc.);
- providing security (to guarantee confidentiality and data integrity).

In general, global trends are such that cloud services are replacing classical architecture of information systems. Therefore we should be prepared for the transfer of existing systems to the cloud [6].

In the cloud technology, a duplication of program code is used to ensure reliability of data transmission. In case of container technologies, only code libraries are duplicated. Some container technology allows avoiding such duplication of system libraries code. The main negative effect of code duplication effect is the heavy load on the CPU cache. The report [7] carries out a detailed analysis of this effect. In practice, the increase of the processor cache use leads to a significant drop of system performance.

One of the features of virtual machines - the lack of direct access to physical memory of the main system. This is one of the reasons that hinder interoperability of the system with input-output devices [8].

For large data storage it is common to split data into smaller segments, which reduces the average access time. This acceleration is achieved by reducing the required number of read operations, during each of them a continuous segment of data can be read from the disk. In all these systems, data storage uses a local file system or a relational database, which significantly limits the possibilities of scaling. When using cloud computing alternative for read operation is establishing a connection and accessing the object in cloud storage. In cloud systems use semi-structured data, hierarchical models, noSQL system and others. Many researchers made it advisable to use graph data representation models [9, 10]. An important feature of the graph methods is the existence of a significant number of polynomial solutions.

For modern cloud applications data transmission networks is a bottleneck. And one of the most important tasks in ensuring the functioning of the systems - configuration of the load in networks, to provide the use of applications in the cloud.

An algorithm for constructing a module of software configuration developed based on dynamic models [11, 12] The review [13] describes the basic QoS assurance system at the network level.

The cloud system, under the conditions of availability of computing resources on the server, there is a static load balancing, in which the distribution is carried out in advance. In the conditions of computing resources constraints and an ever-changing number of requests in the system, static load balancing does not give effect for heavy loaded systems, requiring dynamic load balancing.

Hypervisors emulate almost all the equipment, creating their virtual copies. Parameters of the virtual processor copies, memory, disk, and network adapter, affect the computing performance. Managing this equipment, you can change the virtual machine performance.

Virtual resources are considered to be infinite, but in practice there are physically limited resources on which hypervisor runs. Therefore, the optimal distribution of the physical computing resources between the virtual machines is an important task.

For each virtual machine, the required amount of computing resources is allocated – RAM, the number and frequency of processors, etc. Those are so-called configuration settings of virtual machine. This virtual machines are not always hold all the power selected for their computing, sometimes they are idle, when there are no requests, so at this time, these resources can be used by other virtual machines. This configuration in the hypervisor VMWare ESXi includes a number of parameters, such as Limit, Reservation and Shares for the Resource Pool within the VMware DRS cluster and ESX hosts. It is these three parameters that determine the memory consumption by the virtual machines and CPU resources of VMware ESX host allocated to them.

## II. FORMULATION OF THE PROBLEM

*Limit* determines the limit of consumption of physical resources by the virtual machine pool. Thus, if this parameter is set within the physical resources, there are no conflicts occurred between the virtual machines, but if this parameter is the same for some of the virtual machines, then conflicts may arise in heavy loaded systems. If both processors require more resources than was allocated by the physical parameter Limit, they just start to wait when one processor is released, thus time delays occure.

If memory usage reaches the parameter Limit, then it goes to the *swap* area, which, of course, slower, also causing delays.

The *Shares* parameter defines the prioritization of consumption by the virtual machines among each other within the ESX host and the resource pool. The three standard settings High, Medium and Low priority indicate the priority ratio.

*Reservation* parameter for a virtual machine or resource pool determines how much physical memory or CPU resources will be guaranteed to the virtual machine during operation. If the virtual machine has not yet reached Reservation parameter, the unused resources may be granted to other virtual machines (in Shares); otherwise, if it has already reached that level, then its computing resources will no longer be re-allocated.

Optimally configuring and managing these parameters of computational resources allocation between virtual machines, you can optimize the use of the physical resources of the server by reducing their downtime.

Dynamic load balancing is mainly divided into several sub-tasks: initial allocation of resources; evaluation download compute nodes; initiation of load balancing; taking decisions about balancing; moving objects (migration).

Thus, the problem lies in the fact that the resources of cloud computing environment can perform all the current load with minimal loss of performance. To dynamically manage cloud infrastructure it was suggested to use feedback. For the cloud technology researchers suggest different ways of constructing a system of dynamic equations. Providing a feedback is possible by hypervisors on which modern cloud systems are built, and application software, if such opportunities were provided by developers.

The hypervisor acts as an object that generates a signal that contains the parameters for monitoring the workload of processors, network server's memory and each virtual machine individually. The control mechanism is a software hypervisor management tool, which is based on one or more specified actions that determine the law (algorithm) of management. It generates a control action for the server and maintains a predetermined level or changes the state according to a certain law, which can be displayed on the corresponding output signal.

In almost all applications logging functionality is always provided, through which you can get the time taken by operations. If cross-platform standardized service syslog is used for logging, then with its help you can also send information to the application to monitor and manage the hypervisor (Fig. 1).

As a result, it is possible to display all load parameters for computing resources and query processing time, which allows detecting complex queries to redirect them to the available virtual machines.

A number of studies on the establishment of management systems based on non-linear models with control was conducted. As a controlled parameter process performance ratio is used depending on specified priorities. These parameters have been proposed as promising approaches for achieving QoS with unpredictable load conditions. The purpose was to control the allocation of resources so that the current balance of performance is in line with the priority given to relations under the given constraints. Also, a similar system is implemented to control traffic.

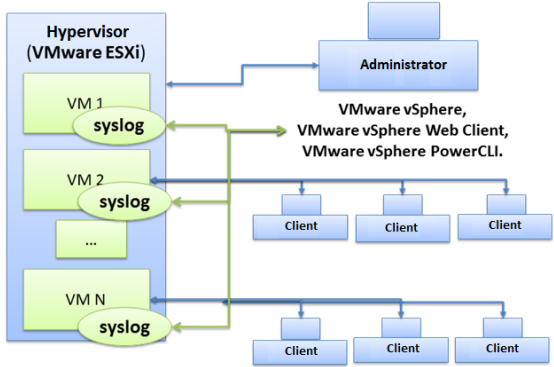

Fig. 1. The infrastructure for monitoring and managing the hypervisor

It should be borne in mind that the introduction of feedback allows to make corrective actions, but at the same time occupies part of computing resources and communication channels, so that is necessary for optimal control problems, where the quality criteria can minimize the processing time and resource constraints. Based on studies on the experimental stand, situations often occur, in which used control method can provide a guaranteed quality of service. These situations include the following: the addition of a new service in the software, which has become very popular, high growth in data volume, need to increase the number of virtual machines, the growing number of geographically dispersed users who require additional cache servers.

### III. SIMULATION

The cloud system, under the conditions of availability of computing resources on the server, there is a static load balancing, in which the distribution is carried out in advance. For this distribution the experience of previous systems and test results are often taken into account. But in the conditions of computing resources constraints and an ever-changing number of requests in the system, static load balancing does not give effect for heavy loaded systems, requiring dynamic load balancing. Dynamic load balancing is essentially divided into several sub-tasks [12, 14, 15]:

- The initial allocation of resources;
- Evaluation of compute nodes load;
- Initiation of load balancing;
- Decision-making about balancing;
- Moving objects (migration).

The problem is in providing enough resources of cloud computing environment ($S_{обл}$) to perform all the current system load with minimal loss of productivity [5]. To solve this problem, monitoring of the system is used, based on which the management is organized.

The formal criterion can be written as:

$$S_{cl} = F(L_{mem}, L_D, L_{cp}, L_{nw}, T_D) \rightarrow max \quad (1)$$

Here $L_{mem}$ is the memory resource, $L_D$ is the drive resource, $L_{cp}$ is the CPU resource, $L_{nw}$ is the network resource, $T_D$ response from drive system.

To assess the resources at cold start of virtual machines in the cloud system, streams are generated to simulate the users' requests with a given intensity.

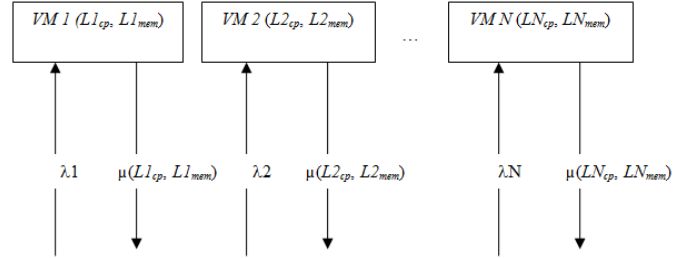

Fig. 2. A simulation model for the processing of cloud system applications

Fig. 2 shows a server with CPU frequency $L_{cp}$ and RAM size of $L_{mem}$, i.e. Lцп and Lоп are limitations of a physical server. Hypervisor is installed on the server with deployed virtual machines, that perform various tasks of processing user requests with incoming intensities λ1, λ2, .., λN. Time of processing the user requests depends on the resources, allocated to virtual machines VM 1 ($L1_{cp}$, $L1_{mem}$), VM 2 ($L2_{cp}$, $L2_{mem}$), ... VM N ($LN_{cp}$, $LN_{mem}$). It also provides the desired level of service. It turns out, the intensities of query processing for users μ(L1цп, L1оп), μ(L2цп, L2оп), ..., μ(LNцп, LNоп) are functions, dependent on allocated for virtual machine CPU and memory resources. It is possible to get empirically the function of the intensity of processing user requests through a series of experiments, timing the execution time of queries with different configurations, and then calculating the μ by the formula:

$$\mu = \frac{1}{\bar{t}}. \quad (2)$$

where $\bar{t}$ is the average time of processing a user query.

The main purpose of the cloud system is to ensure the required quality of service, and thus to ensure the minimum time to process the user queries ($t_q$):

$$\sum_{i=1}^{N} \bar{t}_{iq} \rightarrow min. \quad (2)$$

Since $\bar{t}_q = \frac{\bar{r}}{\lambda}$, where $\bar{r}$ is the average number or queries in a queue, then it should be minimal as well:

$$\sum_{i=1}^{N} \bar{r}_i \rightarrow min \quad (3)$$

If we consider each virtual machine as a single-channel Queueing System (QS) with endless queues, the average number of requests in the queue is calculated as follows: [11]

$$\bar{r} = \frac{\rho^2}{1-\rho}, \text{ where } \rho = \frac{\lambda}{\mu(LN_{cp}, LN_{mem})} < 1. \quad (4)$$

The model of QS depends on problems solved in each virtual machine. If requests can be processed in parallel on a virtual machine, the system should be considered as a multi-channel QS with infinite queue, where the number of channels is determined by the amount of processors allocated to the virtual machine. If processing the query involves all the kernels, the system is a single-channel QS, but the function of processing applications will get another dependency, which is the number of processors.

In the QS with infinite queue, very important condition (4) would be the intensity of the receipt of applications, that has to be less than the intensity of the processing of applications. Otherwise, queue will grow indefinitely. Therefore, when choosing LNцп and LNоп parameters, it determines the minimum search threshold of optimal solutions for (2) and (3), therefore solution obtained in (4) represents the parameters of the Reservation.

When computing resources of the server are enough to perform the tasks

$$\sum_{i=1}^{N} L_{icp} < L_{cp} \quad \text{и} \quad \sum_{i=1}^{N} L_{imem} < L_{mem}, \qquad (5)$$

remaining resources can be identified in the Share for all virtual machines, which will automatically perform the balancing of the system by a hypervisor, increasing the average time of processing user requests.

$$\text{Share}_{cp} = L_{cp} - \sum_{i=1}^{N} L_{icp}, \quad \text{Share}_{озу} = L_{оп} - \sum_{i=1}^{N} L_{iоп}. \quad (6)$$

When computing resources are not enough to perform the tasks and provide the required quality of service to users

$$\sum_{i=1}^{N} L_{icp} > L_{cp} \quad \text{and} \quad \sum_{i=1}^{N} L_{imem} > L_{mem}, \qquad (7)$$

it is necessary to apply the methods of dynamic cloud management system.

## IV. EXPERIMENTS AND RESULTS

Experiments were carried out with the help of the developed software, aimed to control computing resources and run-time operations on a computing server with multiple virtual machines, on which loading applications are installed.

The software on the administrator's workstation performs centralized collection of characteristics of virtual machines, the hypervisor, in addition to time required for processing the system operations. Displaying all of this information on a chronological schedule will allow the administrator to decide on a cloud infrastructure management, as shown in Fig. 3 (red - the CPU usage for VM1; lilac – RAM usage for VM1; blue - CPU usage for VM2; green – RAM usage for VM2; crimson - runtime of the function 1; yellow – runtime of the function 2).

To show all the parameters on one chart it is necessary to normalize them. For the convenience, the parameters are divided into 2 types:

*a)* Computing parameters of virtual machines on the right y-axis, which are displayed as a percentage of the maximum values of the virtual machines.

*b)* Parameters of time required to perform various functions on the left y-axis. Each point of time run-time parameter is marked at the time of the completion of the function, but knowing its time, you can calculate its operating range. Orange in Fig. 3 is a range of work completed in a single function finished at 3:44:10 that runs 25 seconds, respectively, the implementation of which began in 3:43:45. This graph allows the operator to identify more complex queries or transactions, and redirect or move applications to available virtual machines.

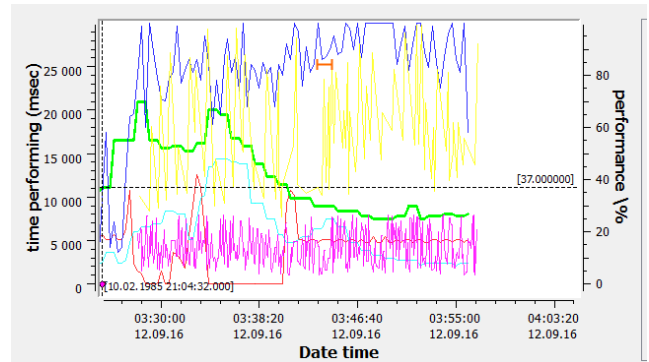

Fig. 3. Graphical display of execution time of functions and computational parameters of virtual machines

The developed method allows determining the display load unbalance moment and deciding on the load balancer. To demonstrate this, the following experiment was conducted.

The application was developed that emulates the load. It performs a recursive search for a file on disk. This application creates separate threads to search at random intervals within a predetermined range, and then closes them at random intervals. The search is creates a heavy load on the CPU computing resources, so other computing parameters will be ignored.

For the experiments a server HP ProLian ML 110 G6 with Intel® processor Xeon® CPU X3450 @ 2.67 Ghz and 4GB RAM was used, with hypervisor VMware ESXi and installed virtual machines running Ubuntu.

The experimental results are shown in Fig. 4, where the blue - CPU usage for VM1, lilac - RAM usage for VM1, red - RAM usage for VM2, green - CPU usage for VM2, raspberry - run-time function f1, which is run from 1 to 10 seconds, and carried out from 1 to 2 seconds, yellow - run time function f2, which is run from 10 to 60 seconds, and was carried out from 5 to 15 seconds, gray - f3 runtime function that starts from 30 to 120 seconds, and was performed for 60 seconds.

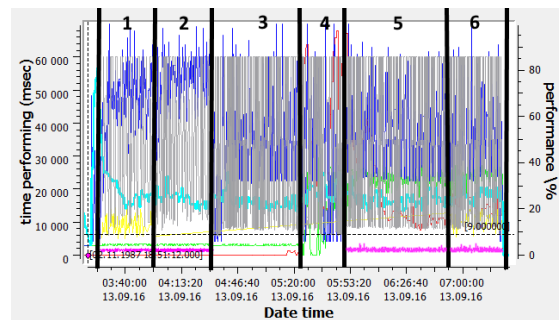

Fig. 4. The experimental results

The whole experience can be divided into six stages.

Two cores and 2GB of RAM are allocated to Virtual machine 1, all three functions are running on it. More results of the phase 1 are shown in Fig. 5. The average CPU usage was in the range of 60-90 percent, which could adversely affect the required quality of service to users. Hence there is a problem, especially for labor-intensive operations, to identify and migrate them to other virtual machines in order to reduce CPU usage.

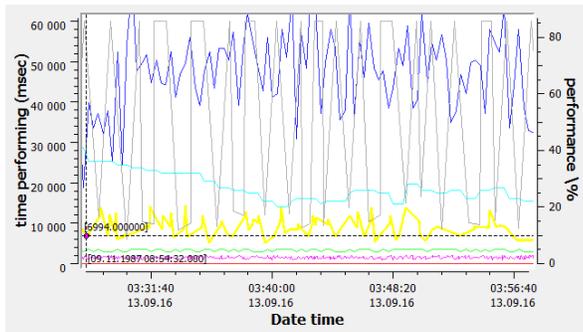

Fig. 5. Phase 1 experimental results

In the second phase, the results of which are shown in Fig. 6, function f2 has been switched off, which, as seen, has no effect on processor usage.

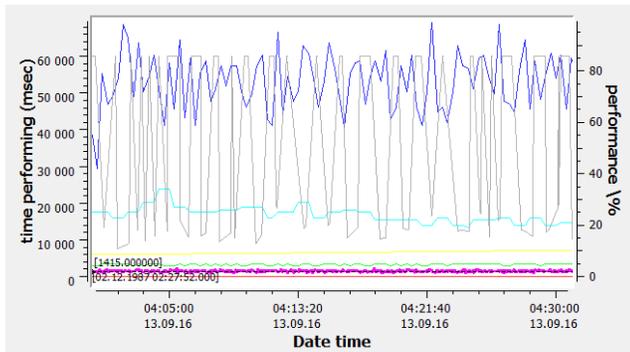

Fig. 6. Phase 2 experimental results

In the third phase, the results of which are shown in Fig. 7, function f1 has been switched of. It did reduce the usage of the processor in the range of 40-80 percent, this leads to the conclusion that the function f3 is the most labor-intensive.

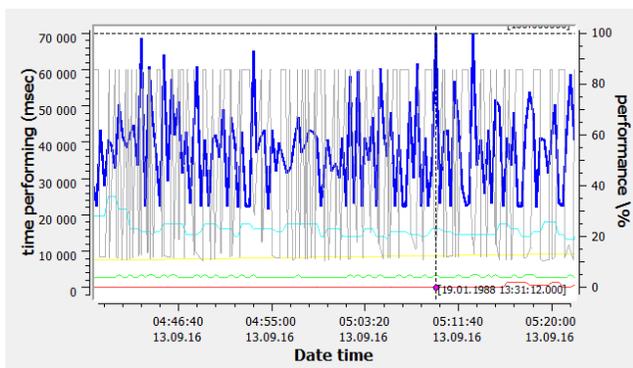

Fig. 7. Phase 3 experimental results

In phase 4 resources of virtual machine 1 were reallocated to another virtual machine, resulting in two identical virtual machines with a single processor core and 1GB of RAM.

In the fifth phase, the results of which are shown in Fig. 8, f3 function remained to work in virtual machine 1, and f1 function, was launched on virtual machine 2. And compared to phase 2 the gain in productivity was obtained in the first virtual machine.

In the last sixth phase, shown in Figure 9, the function f2 has been added to virtual machine number 2, which does not significantly impact the CPU usage on the second virtual machine.

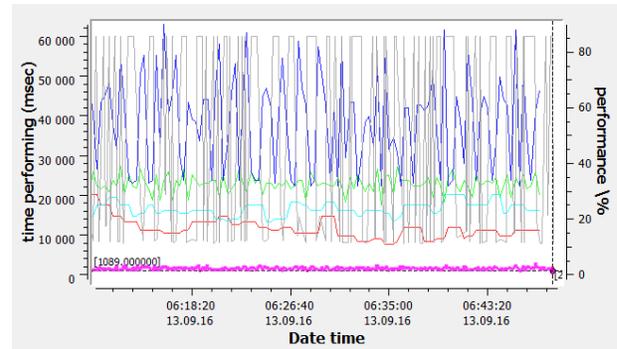

Fig. 8. Phase 5 experimental results

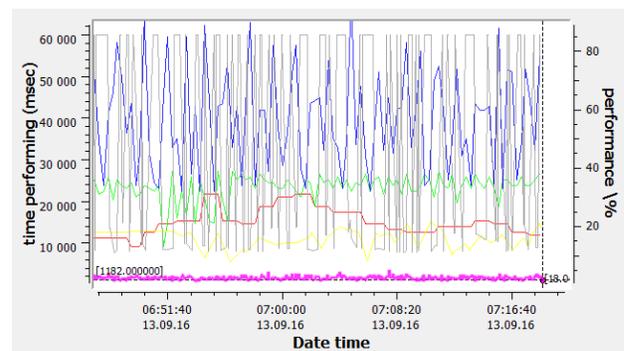

Fig. 9. Phase 6 experimental results

As a result, using the same amount of computing resources, insignificant gains were obtained in the performance of a virtual machine 1 running the function f3, while other functions f1 and f2 were isolated, thus providing the desired quality of service for at least these functions . If in Phase 1 f3 affected the processing of functions f1 and f2, then by isolating them makes it impossible, which has the positive effect on the required quality of service for users.

## V. DISCUSSION

Too frequent load balancing can lead to the case, where simulation model slows down. The costs of balancing itself may surpass the possible benefit from its implementation. Therefore, for productive balancing it is necessary to determine the time of its initialization.

It requires:

- Determine the time of the load imbalance.
- Determine the degree of needed balancing by comparing the potential benefits of its implementation and the cost of it.

Load imbalance can be determined synchronously and asynchronously.

In synchronous determination of imbalance, all processors (network computers) are interrupted at certain times of synchronization, and imbalance is determined by comparing the load on a separate processor with a total average load. In asynchronous determination of imbalance, each processor keeps a history of its usage. In this case, the moment of synchronization for determination of imbalance is absent. The background process that runs in parallel with the application calculates the imbalance.

Most of the dynamic load-balancing strategies can be classified as centralized or fully distributed. With centralized strategy, one computer collects global information on the status of the entire computer system and makes a decision about moving tasks between the computers. With fully distributed strategy, each processor performs load-balancing algorithm to exchange information on the status with other processors. Migration occurs only between neighboring processors.

## VI. Conclusion

The article proposed a method of initial allocation of computing resources for the virtual machines in the hypervisor using a simulation model of the processing user requests, which solved the problems (2) and (3) under the conditions (4-6). The article also lists the software for collection and analysis of computational load of the virtual machines in the hypervisor and the time of the effectiveness of the objective functions of applications that handle user requests. Based on this data, the operator has the possibility to reallocate the computational resources of the hypervisor, thereby providing dynamic control for the system.

The experiment showed the effectiveness of the use of this application. The most labor-intensive function was found, which, in consequence, was isolated from the other functions, which ensured their required level of service. The above methods for configuring and managing hypervisors allow better use of computing resources of servers.

In the long term development of the use of these methods it is possible to develop a algorithm that performs automatic migration of services, or redirecting requests that will organize the distribution of computing in the misty system.

## Acknowledgment

This work was partially funded by the basic part of state task of the Ministry of education and science of Russia for scientific project #792: Research and development of methods and algorithms for constructing open distributed information systems for various purposes. The work is funded by the Moscow Institute of Technology.